\documentclass[letterpaper, 10 pt, conference]{ieeeconf}  % Comment this line out if you need a4paper

\IEEEoverridecommandlockouts                              % This command is only needed if 
\usepackage{graphics} % for pdf, bitmapped graphics files
\usepackage[pdftex]{graphicx}
\usepackage{amsmath} % assumes amsmath package installed
\usepackage{amssymb}  % assumes amsmath package installed
\usepackage{cite}
\usepackage[ruled,vlined]{algorithm2e}
\usepackage{tikz}

\newcommand{\rpartial}{{\mathrm{\partial}}}

\newcommand{\rd}{{\mathrm d}}

\newcommand{\rtr}{{\mathrm{tr}}}

\newcommand{\rT}{{\mathrm{T}}}

\newcommand{\E}{{\mathbb{E}}}

\newcommand{\vx}{{\bf x}}

\newcommand{\vz}{{\bf z}}
\newcommand{\vp}{{\bf p}}

\newcommand{\vf}{{\bf f}}

\newcommand{\vG}{{\bf G}}
\newcommand{\vu}{{\bf u}}

\newcommand{\vw}{{\bf w}}

\newcommand{\vR}{{\bf R}}

\newcommand{\vH}{{\bf H}}

\newcommand{\vB}{{\bf B}}

\newcommand{\vcalG}{{\mbox{\boldmath$\cal{G}$}}}

\newcommand{\vcalH}{{\mbox{\boldmath$\cal{H}$}}}

\newcommand{\p}{\mathsf{p}}
\newcommand{\q}{\mathsf{q}}

\newcommand{\Qb}{\mathbb{Q}}
\newcommand{\Pb}{\mathbb{P}}

\newcommand{\ExP}[2]{\E_{{#1}}{\left[#2\right]}}

\usepackage{amsmath}
\usepackage{amssymb}
\usepackage{xspace}
\usepackage{breqn} % for breaking lines in equations that are too long
\newtheorem{theorem}{Theorem}
\title{\LARGE \bf Model Predictive Path Integral Control using Covariance Variable Importance Sampling
}

\author{Grady Williams{$^1$}, Andrew Aldrich$^1$, and Evangelos A. Theodorou{$^1$} % <-this % stops a space
\thanks{This research has been supported by NSF Grant No. NRI-1426945. The ${^1}$authors are with the Autonomous Control and Decision Systems Laboratory at the Georgia Institute of Technology, 
Atlanta, 
GA, USA. 
Email: gradyrw@gatech.edu}% <-this % stops a space
}

\begin{document}

\maketitle
\thispagestyle{empty}
\pagestyle{empty}

%%%%%%%%%%%%%%%%%%%%%%%%%%%%%%%%%%%%%%%%%%%%%%%%%%%%%%%%%%%%%%%%%%%%%%%%%%%%%%%%
\begin{abstract}
In this paper we develop a Model Predictive Path Integral (MPPI) control algorithm based on a generalized importance sampling scheme and perform parallel optimization via sampling  using a Graphics Processing Unit (GPU).  The proposed generalized importance sampling scheme allows for changes in the drift and diffusion terms of stochastic diffusion processes and plays a significant role in the performance of the model predictive control algorithm. We compare the proposed algorithm in simulation with a model predictive control version of differential dynamic programming.
\end{abstract}

%%%%%%%%%%%%%%%%%%%%%%%%%%%%%%%%%%%%%%%%%%%%%%%%%%%%%%%%%%%%%%%%%%%%%%%%%%%%%%%%
\section{INTRODUCTION}

The path integral optimal control framework \cite{TheodorouCDC2012, Kappen1995, theodorou2015_Entropy} provides a mathematically sound methodology for developing optimal control algorithms based on stochastic sampling of trajectories. The key idea in this framework is that the value function for the optimal control problem is transformed using the Feynman-Kac lemma \cite{Karatzasbook,Friedman75} into an expectation over all possible trajectories, which is known as a path integral.  This transformation allows stochastic optimal control problems to be solved with a Monte-Carlo approximation using forward sampling of stochastic diffusion processes. 

There have been a variety of algorithms developed in the path integral control setting. The most straight-forward  application of path integral control is  when the iterative feedback control law suggested in  \cite{TheodorouCDC2012} is implemented in its open loop formulation.  This requires that sampling  takes place only from the initial state of the optimal control problem. A more effective approach is to use the path integral control framework to find the parameters of a feedback control policy. This can be done by sampling in policy parameter space, these methods are known as Policy Improvement with Path Integrals \cite{theodorou2010}. Another approach to finding the parameters of a policy is to attempt to directly sample from the optimal distribution defined by the value function \cite{gomez2014policy}. Other methods along similar threads of research include \cite{thijssen2015path, Rombokas2013}. 

Another way that the path integral control framework can be applied is in a model predictive control setting. In this setting an open-loop control sequence is constantly optimized in the background while the machine is simultaneously executing the ``best guess'' that the controller has. An issue with this approach is that many trajectories must be sampled in real-time, which is difficult when the system has complex dynamics. One way around this problem is to drastically simplify the system under consideration by using a hierarchical scheme \cite{gomez2015real}, and use path integral control to generate trajectories for a point mass which is then followed by a low level controller. Even though this approach may be successfull for certain applications, it is limited in the kinds of behaviors that it can generate since it does not consider the full non-linearity of dynamics.     A more efficient approach is  to take advantage of the parallel nature of sampling and use a graphics processing unit (GPU) \cite{Willams2014} to sample thousands of trajectories from the nonlinear dynamics.  

A major issue in the path integral control framework is that the expectation is taken with respect to the uncontrolled  dynamics of the system. This is problematic since the probability of sampling a low cost trajectory using the uncontrolled dynamics is typically very low. This problem becomes more  drastic when the underlying dynamics are nonlinear and sampled trajectories can become trapped in undesirable  parts of the state space.  It has previously been demonstrated how to change the mean of the sampling distribution using Girsanov's theorem \cite{theodorou2015_Entropy,TheodorouCDC2012}, this can then be used to develop an iterative algorithm. However, the variance of the sampling distribution has always remained unchanged. Although in some simple simulated scenarios changing the variance is not necessary, in many cases the natural variance of a system will be too low to produce useful deviations from the current trajectory. Previous methods have either dealt with this problem by artificially adding noise into the system and then optimizing the noisy system \cite{Rombokas2013, theodorou2010}. Or they have simply ignored the problem entirely and sampled from whatever distribution worked best \cite{stulp10reinforcement, Willams2014}. Although these approaches can be successful, both are problematic in that the optimization either takes place with respect to the wrong system or the resulting algorithm ignores the theoretical basis of path integral control.

The approach we take here generalizes these approaches in that it enables for both the mean and variance of the sampling distribution to be changed by the control designer, without violating the underlying assumptions made in the path integral derivation. This enables the algorithm to converge fast enough that it can be applied in a model predictive control setting. After deriving the model predictive path integral control (MPPI) algorithm, we compare it with an existing model predictive control formulation based on differential dynamic programming (DDP) \cite{jacobson1970,EvangelosACC2010,Todorov2005b}. DDP is one of the most powerful techniques for trajectory optimization, it relies on a first or second order approximation of the dynamics and a quadratic approximation of the cost along a nominal trajectory, it then computes a second order approximation of the value function which it uses to generate the control.

\section{PATH INTEGRAL CONTROL}

 In this section we review the path integral optimal control framework \cite{Kappen1995}. Let $\vx_t \in \mathbb{R}^N$ denote the state of a dynamical system at time $t$, $\vu(\vx_t, t) \in \mathbb{R}^m$ denotes a control input for the system, $\tau: [t_0, T] \rightarrow \mathbb{R}^n$ represents a trajectory of the system, and $\rd \vw \in \mathbb{R}^p$ is a brownian disturbance. In the path integral control framework we suppose that the dynamics take the form:
\begin{equation}
\label{Equation:SystemDynamics}
\rd \vx = \vf(\vx_t,t)\rd t + \vG(\vx_t,t)\vu(\vx_t,t) \rd t + \vB(\vx_t,t)\rd \vw 
\end{equation}
In other words, the dynamics are affine in control and subject to an affine brownian disturbance. We also assume that $\vG$ and $\vB$ are partitioned as:
\begin{equation}
\label{Equation:PartitionedSystem}
\vG(\vx_t, t) = \begin{pmatrix} 0 \\ \vG_c(\vx_t, t) \end{pmatrix}; ~~ \vB(\vx_t, t) = \begin{pmatrix} 0 \\ \vB_c(\vx_t, t) \end{pmatrix}
\end{equation}
Expectations taken with respect to (\ref{Equation:SystemDynamics}) are denoted as $\ExP{\Qb}{\cdot}$, we will also be interested in taking expectations with respect to the uncontrolled dynamics of the system (i.e (\ref{Equation:SystemDynamics}) with $\vu \equiv 0$). These will be denoted $\ExP{\Pb}{\cdot}$. We suppose that the cost function for the optimal control problem has a quadratic control cost and an arbitrary state-dependent cost. Let $\phi(\vx_{T})$ denote a final the terminal cost,  $q(\vx_t, t) $ a state dependent running cost, and define $\vR(\vx_t, t)$ as a positive definite matrix. The value function $V(\vx_t, t)$ for this optimal control problem is then defined as:
\begin{equation}
\label{Equation:ValueFunction}
\min_{\vu} \ExP{\Qb}{ \phi(\vx_{T}) + \int_{t}^{T}  \bigg(q(\vx_t,t) + \frac{1}{2} \vu^\rT \vR(\vx_t, t) \vu \bigg)\rd t  } 
\end{equation}
The Stochastic Hamilton-Jacobi-Bellman equation \cite{Stengel1994,Fleming2006} for the type of system in \eqref{Equation:SystemDynamics}  and for  the cost function in \eqref{Equation:ValueFunction} is given as:
\begin{dmath}
\label{Equation:StochasticHJBU4}
-\rpartial_tV = q(\vx_t, t) + \vf(\vx_t, t)^\rT V_\vx - \frac{1}{2} V_\vx^\rT \vG(\vx_t, t) \vR(\vx_t, t)^{-1} \vG(\vx_t, t)^\rT V_\vx + \frac{1}{2}\rtr(\vB(\vx_t, t) \vB(\vx_t, t)^\rT V_{\vx \vx})
\end{dmath}
where the optimal control is expressed as:  
\begin{equation}
\vu^* = -\vR(\vx_t, t)^{-1} \vG(\vx_t, t)^\rT V_\vx
\end{equation}
The solution to this backwards PDE yields the value function for the stochastic optimal control problem, which is then used to generate the optimal control. Unfortunately, classical methods for solving partial differential equations of this nature suffer from the curse of dimensionality and are intractable for systems with more than a few state variables. 

The approach we take in the path integral control framework is to transform the backwards PDE into a path integral, which is an expectation over all possible trajectories of the system. This expectation can then be approximated by forward sampling of the stochastic dynamics. In order to effect this transformation we apply an exponential transformation of the value function 
\begin{equation}
V(\vx, t) = -\lambda \log(\Psi(\vx, t))
\end{equation}
Here $\lambda$ is a positive constant. We also have to assume a relationship between the cost and noise in the system (as well as $\lambda$) through the equation:
\begin{equation}
\vB_c(\vx_t, t) \vB_c(\vx_,t)^\rT = \lambda \vG_c(\vx_t, t) \vR(\vx_t, t)^{-1} \vG_c(\vx_t, t)^\rT
\end{equation}
The main restriction implied by this assumption is that $\vB(\vx_t, t)$ has the same rank as $\vR(\vx_t, t)$. This limits the noise in the system to only effect state variables that are directly actuated (i.e. the noise is control dependent). There are a wide variety of systems which naturally fall into this description, so the assumption is not too restrictive. However, there are interesting systems for which this description does not hold (i.e. if there are known strong disturbances on indirectly actuated state variables or if the dynamics are only partially known). 

By making this assumption and performing the exponential transformation of the value function the stochastic HJB equation is transformed into the \emph{linear} partial differential equation:
\begin{equation}
\label{Equation:LinearizedHJB}
\rpartial_t \Psi = \frac{\Psi(\vx_t, t)}{\lambda}q(\vx_t, t) - \vf(\vx_t, t)^\rT \Psi_\vx  - \frac{1}{2} \rtr(\Sigma(\vx_t, t) \Psi_{\vx \vx})
\end{equation}
Here we've denoted the covariance matrix $$\vB_c(\vx_t, t) \vB_c(\vx_t, t)^\rT$$ as $\Sigma(\vx_t, t)$.
This equation is known as the backward Chapman-Kolmogorov PDE. We can then apply the Feynman-Kac lemma, which relates backward PDEs of this type to path integrals through the equation:
\begin{equation}
\label{Equation:FeynmanKac}
\Psi(\vx_{t_0}, t_0) = \ExP{\Pb}{ \exp\left(-\frac{1}{\lambda} \int_{t_0}^T q(\vx, t) ~ \rd t \right) \Psi(\vx_{T}, T) }
\end{equation}
Note that the expectation (which is the path integral) is taken with respect to $\Pb$ which is the uncontrolled dynamics of the system. By recognizing that the term $\Psi(\vx_T)$ is the transformed terminal cost: $e^{-\frac{1}{\lambda} \phi(\vx_T)}$ we can re-write this expression as:
\begin{equation}
\label{Equation:PathIntegralDiscrete}
\Psi(\vx_{t_0}, t_{0}) \approx \ExP{\Pb}{ \exp\left(-\frac{1}{\lambda} S(\tau) \right)}
\end{equation}
where $S(\tau) = \phi(\vx_T) + \int_{t_0}^T q(\vx_t, t) \rd t $ is the cost-to-go of the state dependent cost of a trajectory. Lastly we have to compute the gradient of $\Psi$ with respect to the initial state $\vx_{t_0}$. This can be done analytically and is a straightforward, albeit lengthy, computation so we omit it and refer the interested reader to \cite{theodorou2010}. After taking the gradient we obtain:
\begin{dmath}
\label{Equation:PathIntegralControlFramework}
\vu^* \rd t = \vcalG(\vx_{t_0}, t_0)^{-1} \frac{\ExP{\Pb}{ \exp \left(-\frac{1}{\lambda}S(\tau) \right) \vB(\vx_{t_0}, t_0) \rd \vw}}{\ExP{\Pb}{\ \exp \left(-\frac{1}{\lambda}S(\tau) \right) }}
\end{dmath}
Where the matrix $\vcalG(\vx_t, t)$ is defined as:
\begin{equation}
\vR(\vx_t, t)^{-1} \vG_c(\vx_t, t)^\rT \left( \vG_c(\vx_t,t) \vR(\vx_t, t)^{-1} \vG_c(\vx_t, t)^\rT \right)^{-1}
\end{equation} 
Note that if $\vG_c(\vx_t, t)$ is square (which is the case if the system is not over actuated) this reduces to $\vG_c(\vx_t, t)^{-1}$.

Equation (\ref{Equation:PathIntegralControlFramework}) is the path integral form of the optimal control. The fundamental difference between this form of the optimal control and classical optimal control theory is that instead of relying on a backwards in time process, this formula requires the evaluation of an expectation which can be approximated using forward sampling of stochastic differential equations.

\subsection{Discrete Approximation}

Equation (\ref{Equation:PathIntegralControlFramework}) provides an expression for the optimal control in terms of a path integral. However, these equations are for continuous time and in order to sample trajectories on a computer we need discrete time approximations. 

We first discretize the dynamics of the system. We have that $\vx_{t+1} = \vx_t + \rd \vx_t$ where $\rd \vx_t$ is defined as:
\begin{equation}
\rd \vx_t = \left(\vf(\vx_t,t) + \vG(\vx_t,t)\vu(\vx_t,t)\right) \Delta t+ \vB(\vx_t,t)\epsilon \sqrt{\Delta t}
\end{equation}
The term $\epsilon$ is a vector of standard normal Gaussian random variables. For the uncontrolled dynamics of the system we have:
\begin{equation}
\rd \vx_t = \vf(\vx_t,t)\Delta t + \vB(\vx_t,t)\epsilon \sqrt{\Delta t}
\end{equation}
Another way we can express $\vB(\vx_t, t) \rd \vw$ which will be useful is as:
\begin{equation}
\vB(\vx_t, t)\rd \vw \approx \rd \vx_t - \vf(\vx_t, t)\Delta t
\end{equation}
Lastly we say: $S(\tau) \approx \phi(\vx_T) + \sum_{i=0}^N q(\vx_t, t) \Delta t$ where $N = (T - t)/\Delta t$ Then by defining $\p$ as the probability induced by the discrete time uncontrolled dynamics we can approximate (\ref{Equation:PathIntegralControlFramework}) as:
\begin{equation}
\label{Equation:DiscretePI}
\vu^* = \vcalG(\vx_{t_0}, t_0)^{-1} \frac{\ExP{\vp}{ \exp \left(-\frac{1}{\lambda}S(\tau) \right) \left( \frac{\rd \vx_{t_0}}{\Delta t} - \vf(\vx_{t_0}, t_0) \right)}}{\ExP{\vp}{\ \exp \left(-\frac{1}{\lambda}S(\tau) \right) }}
\end{equation}
Note that we have moved the $\Delta t$ term multiplying $\vu$ over to the right-hand side of the equation and inserted it into the expectation.

\section{GENERALIZED IMPORTANCE SAMPLING}

Equation (\ref{Equation:DiscretePI}) provides an implementable method for approximating the optimal control via random sampling of trajectories. By drawing many samples from $\p$ the expectation can be evaluated using a Monte-Carlo approximation. In practice, this approach is unlikely to succeed. The problem is that $\p$ is typically an inefficient distribution to sample from (i.e the cost-to-go will be high for most trajectores sampled from $\p$). Intuitively sampling from the uncontrolled dynamics corresponds to turning a machine on and waiting for the natural noise in the system dynamics to produce interesting behavior. 

In order to efficiently approximate the controls, we require the ability to sample from a distribution which is likely to produce low cost trajectories. In previous applications of path integral control \cite{theodorou2015_Entropy, TheodorouCDC2012} the mean of the sampling distribution has been changed which allows for an iterative update law. However, the variance of the sampling distribution has always remained unchanged. In well engineered systems, where the natural variance of the system is very low, changing the mean is insufficient since the state space is never aggressively explored. In the following derivation we provide a method for changing both the initial control input and the variance of the sampling distribution.

\subsection{Likelihood Ratio}

We suppose that we have a sampling distribution with non-zero control input and a changed variance, which we denote as $\q$, and we would like to approximate (\ref{Equation:DiscretePI}) using samples from $\q$ as opposed to $\p$. Now if we write the expectation term (\ref{Equation:DiscretePI}) in integral form we get:
\begin{equation}
\label{Equation:IntegralPI}
\frac{\int \exp \left(-\frac{1}{\lambda}S(\tau) \right) \left( \frac{\rd \vx_{t_0}}{\Delta t} - \vf(\vx_t, t) \right) \p(\tau) \rd \tau}{\int \exp \left(-\frac{1}{\lambda}S(\tau) \right) \p(\tau) \rd \tau}
\end{equation}
Where we are abusing notation and using $\tau$ to represent the discrete trajectory $(\vx_{t_0}, \vx_{t_1}, \dots \vx_{t_N})$. Next we multiply both integrals by $1 = \frac{\q(\tau)}{\q(\tau)}$ to get:
\begin{equation}
\label{Equation:IntegralPI__II}
\frac{\int \exp \left(-\frac{1}{\lambda}S(\tau) \right) \left( \frac{\rd \vx_{t_0}}{\Delta t} - \vf(\vx_t, t) \right) \frac{\q(\tau)}{\q(\tau)} \p(\tau) \rd \tau}{\int \exp \left(-\frac{1}{\lambda}S(\tau) \right)\frac{\q(\tau)}{\q(\tau)} \p(\tau) \rd \tau}
\end{equation}
And we can then write this as an expectation with respect to $\q$:
\begin{equation}
\label{Equation:ImportanceTerm}
\frac{\ExP{\q}{ \exp \left(-\frac{1}{\lambda}S(\tau) \right) \left( \frac{\rd \vx_{t_0}}{\Delta t} - \vf(\vx_t, t) \right) \frac{\p(\tau)}{\q(\tau)}}}{\ExP{\q}{ \exp \left(-\frac{1}{\lambda}S(\tau) \right)\frac{\p(\tau)}{\q(\tau)}}}
\end{equation}
We now have the expectation in terms of a sampling distribution $\q$ for which we can choose:
\begin{enumerate}
\item The initial control sequence from which to sample around.
\item The variance of the exploration noise which determines how aggressively the state space is explored.
\end{enumerate}
However, we now have an extra term to compute $\frac{\p(\tau)}{\q(\tau)}$. This is known as the \emph{likelihood ratio} (or Radon-Nikodym derivative) between the distributions $\p$ and $\q$. In order to derive an expression for this term we first have to derive equations for the probability density functions of $\p(\tau)$ and $\q(\tau)$ individually. We can do this by deriving the probability density function for the general discrete time diffusion processes $P(\tau)$, corresponding to the dynamics: 
\begin{equation}
\rd \vx_t = \left(\vf(\vx_t,t) + \vG(\vx_t,t)\vu(\vx_t,t)\right) \Delta t+ \vB(\vx_t,t)\epsilon \sqrt{\Delta t}
\end{equation}
The goal is to find $P(\tau) = P(\vx_{t_0}, \vx_{t_1}, \dots \vx_{t_N})$. By conditioning and using the Markov property of the state space this probability becomes:
\begin{equation}
P(\vx_{t_0}, \vx_{t_1}, \dots \vx_{t_N}) = \prod_{i=1}^N P(\vx_{t_i} | \vx_{t_{i-1}})
\end{equation}
Now recall that a portion of the state space has deterministic dynamics and that we've partitioned the diffusion matrix as:
\begin{equation}
\vB(\vx_t, t) = \begin{pmatrix}
0 \\ \vB_c(\vx_t, t)
\end{pmatrix}
\end{equation}
We can partition the state variables $\vx$ into the deterministic and non-deterministic variables $\vx_t^{(a)}$ and $\vx_t^{(c)}$ respectively. The next step is to condition on $\vx_{t+1}^{(a)} = F^{(a)}(\vx_t, t) = \vx_t^{(a)} + \left( \vf^{(a)}(\vx_t, t) + \vG^{(a)}(\vx_t, t)\vu_t \right) \rd t $ since if this does not hold $P(\tau)$ is zero. We thus need to compute:
\begin{equation}
\prod_{i=1}^N P \left(\vx_{t_i} |\vx_{t_{i-1}}, \vx_{t_i}^{(a)} = F^{(a)}(\vx_{t_{i-1}}, t_{i-1} \right)
\end{equation}
And from the dynamics equations we know that each of these one-step transitions is Gaussian with mean: $\vf^{(c)}(\vx_t, t) + \vG^{(c)}(\vx_{t_i}, t_i)\vu(\vx_{t_i}, t_i)$ and variance:
\begin{equation}
\Sigma_i = \vB_c(\vx_{t_i}, t_i)\vB_c(\vx_{t_i}, t_i)^\rT \Delta t.
\end{equation}
We then define $\vz_i = \frac{\rd \vx_{t_i}^{(c)}}{\Delta t} - f^{(c)}(\vx_{t_i}, t_i)$, and $\mu_i = \vG^{(c)}(\vx_{t_i}, t_i)\vu(\vx_{t_i}, t_i)$. Applying the definition of the Gaussian distribution with these terms yields:
\begin{equation}
P(\tau) = \prod_{i=1}^N \frac{\exp\left( -\frac{\Delta t}{2} \left(\vz_i - \mu_i \right)^\rT \Sigma_i^{-1}\left(\vz_i - \mu_i \right) \right)}{(2 \pi)^{n/2} |\Sigma_i |^{1/2}}
\end{equation}
And then using basic rules of exponents this probability becomes:
\begin{equation}
\label{Equation:traj_pdf}
Z(\tau)^{-1} \exp \left(-\frac{\Delta t}{2}\sum_{i=1}^N \left(\vz_i - \mu_i \right)^\rT \Sigma_i^{-1}\left(\vz_i - \mu_i \right)\right)
\end{equation}
Where $Z(\tau) = \prod_{i=1}^N (2 \pi)^{n/2} |\Sigma_i |^{1/2}$.  With this equation in hand we're now ready to compute the likelihood ratio between two diffusion processes.
\vspace{2mm}
\begin{theorem}
Let $\p(\tau)$ be the probability density function for trajectories under the uncontrolled discrete time dynamics:
\begin{equation}
\rd \vx_{t} = \vf(\vx_{t}, t)\Delta t + \vB(\vx_{t}, t)\epsilon \sqrt{\Delta t}
\end{equation}
And let $\q(\tau)$ be the probability density function for trajectories under the controlled dynamics
with an adjusted variance:
\begin{multline}
\rd \vx_{t} = \left( \vf(\vx_{t}, t) + \vG(\vx_{t}, t)\vu(\vx_{t}, t) \right)\Delta t + \\ \vB_E(\vx_{t}, t)\epsilon \sqrt{\Delta t}
\end{multline}
Where the adjusted variance has the form: $$\vB_E(\vx_t, t) = \begin{pmatrix} 0 \\ A_t \vB_c(\vx_t, t) \end{pmatrix}$$
And define $\vz_i$, $\mu_i$, and $\Sigma_i$ as before. Let $Q_i$ be defined as:
\begin{dmath}
\label{Equation:zeta}
Q_i = \left(\vz_i - \mu_i \right)^\rT \Gamma_i^{-1} \left(\vz_i - \mu_i \right) + 2\left(\mu_i \right)^\rT \Sigma_i^{-1} \left(\vz_i - \mu_i \right) + \mu_i^\rT \Sigma_i^{-1} \mu_i
\end{dmath}
Where $\Gamma_i$ is:
\begin{equation}
\Gamma_i^{-1} = \left( \Sigma_i^{-1} - A_{t_i}^\rT \Sigma_i A_{t_i} \right)^{-1}
\end{equation}
Then under the condition that each $A_{t_i}$ is invertible and each $\Gamma_i$ is invertible, the likelihood ratio for the two distributions is:
\begin{equation}
\left(\prod_{i=1}^N |A_{t_i}|\right) \exp \left(-\frac{\Delta t}{2} \sum_{i=1}^N Q_i \right)
\end{equation}
\end{theorem} 
\vspace{5mm}
\begin{proof}
In discrete time the probability of a trajectory is formulated according to the (\ref{Equation:traj_pdf}). We thus have $\p(\tau)$ equal to:
\begin{equation}
\p(\tau) = \frac{\exp \left( -\frac{\Delta t}{2} \sum_{i=1}^N \vz_i \Sigma_i \vz_i \right)}{Z_\p(\tau)}
\end{equation}
and $q(\tau)$ equal to:
\begin{equation}
\resizebox{.99\hsize}{!}{$
\frac{\exp \left( -\frac{\Delta t}{2} \sum_{i=1}^N \left(\vz_i - \mu_i \right)^\rT \left(A_{t_i}^\rT\Sigma_i A_{t_i} \right)^{-1} \left(\vz_i - \mu_i \right)\right)}{Z_\q(\tau)}
$}
\end{equation}
Then dividing these two equations we have {\large{$\frac{\p(\tau)}{\q(\tau)}$}} as:
\begin{equation}\label{Equation:Radon_Nikodym}
\left( \prod_{i=1}^N \frac{(2 \pi)^{n/2} |A_{t_i}^\rT \Sigma_i A_{t_i}|^{1/2}) }{(2 \pi)^{n/2} |\Sigma_i|^{1/2}) }\right)\exp \left( -\frac{\Delta t}{2} \sum_{i=1}^N \zeta_i \right)
\end{equation}
Where $\zeta_i$ is:
\begin{dmath}
\zeta_i = \left( \vz_i^\rT \Sigma_i^{-1} \vz_i - (\vz_i - \mu_i)^\rT \left( A_{t_i}^\rT \Sigma_i A_{t_i} \right)^{-1} (\vz_i - \mu_i) \right)
\end{dmath}
Using basic rules of determinants it is easy to see that the term outside the exponent reduces to
\begin{equation} 
 \prod_{j=1}^N \frac{(2 \pi)^{n/2} |A_j^\rT \Sigma_j A_j|^{1/2}) }{(2 \pi)^{n/2} |\Sigma_j|^{1/2}) } = \prod_{j=1}^N |A_j|
\end{equation}
So we need only show that $\zeta_i$ reduces to $Q_i$. Observe that at every timestep we have the difference between two quadratic functions of $\vz_i$, so we can complete the square to combine this into a single quadratic function. If we recall the definition of $\Gamma_i$ from above, and define $\Lambda_i = A_{t_i}^\rT \Sigma_i A_{t_i}$ then completing the square yields:
\begin{dmath}
\zeta_i =  \left( \vz_i + \Gamma_i \Lambda_i^{-1} \mu_i \right)^\rT \Gamma_i^{-1} \left( \vz_i + \Gamma_i \Lambda_i^{-1}\mu_i \right) -
\mu_i^\rT \Lambda_i^{-1}\mu_i -
\left( \Gamma_i \Lambda_i^{-1} \mu_i \right)^\rT \Gamma_i^{-1} \left( \Gamma_t \Lambda_i^{-1} \mu_i \right)
\end{dmath}
Now we expand out the first quadratic term to get:
\begin{dmath}
\zeta_i =  \vz_i^\rT \Gamma_i^{-1} \vz_i + 2\mu_i^\rT \Lambda_i^{-1} \vz_i + \underline{\mu_i^\rT \Lambda_i^{-1} \Gamma_i \Lambda_i^{-1} \mu_i} - \mu_i^\rT \Lambda_i^{-1}\mu_i - \underline{(\Gamma_i \Lambda_i^{-1} \mu_i)^\rT \Gamma_i^{-1} (\Gamma_i \Lambda_i^{-1} \mu_i)}
\end{dmath}
Notice that the two underlined terms are the same, except for the sign, so they cancel out and we're left with:
\begin{dmath}
\zeta_i =  \vz_i^\rT \Gamma_i^{-1} \vz_i + 2\mu_i^\rT \Lambda_i^{-1} \vz_i - \mu_i^\rT \Lambda_i^{-1}\mu_i
\end{dmath}
Now define $\tilde{\vz}_i = \vz_i - \mu_i$, and then re-write this equation in terms of $\tilde{\vz_i}$:
\begin{equation}
\zeta_i =  (\tilde{\vz}_i + \mu_i)^\rT \Gamma_i^{-1} (\tilde{\vz}_i + \mu_i) + 2\mu_i^\rT \Lambda_i^{-1} (\tilde{\vz}_i + \mu_i) - \mu_i^\rT \Lambda_i^{-1}\mu_i
\end{equation}
which expands out to:
\begin{dmath}
\zeta_i = \tilde{\vz}_i^\rT \Gamma_i^{-1} \tilde{\vz}_i + 2\mu_i^\rT \Gamma_i^{-1} \tilde{\vz}_i + \mu_i^\rT \Gamma_i^{-1} \mu_i + 2\mu_i^\rT \Lambda_i^{-1} \tilde{\vz}_i + 2\mu_i^\rT \Lambda_i^{-1} \mu_i - \mu_i^\rT \Lambda_i^{-1}\mu_i
\end{dmath}
Which then simplifies to:
\begin{dmath}
\zeta_i = \tilde{\vz}_i^\rT \Gamma_i^{-1} \tilde{\vz}_i + 2\mu_i^\rT \Gamma_i^{-1} \tilde{\vz}_i + \mu_i^\rT \Gamma_i^{-1} \mu_i + 2\mu_i^\rT \Lambda_i^{-1} \tilde{\vz}_i + \mu_i^\rT \Lambda_i^{-1} \mu_i 
\end{dmath}
Now recall that $\Gamma_i = (\Sigma_i^{-1} - \Lambda_i^{-1})^{-1}$, so we can split the quadratic terms in $\Gamma_i^{-1}$ into the $\Sigma_i^{-1}$ and $\Lambda_i^{-1}$ components. Doing this yields:
\begin{dmath}
\zeta_i = \tilde{\vz}_i^\rT \Gamma_i^{-1} \tilde{\vz}_i + 2\mu_i^\rT \Sigma_i^{-1} \tilde{\vz}_i - \underline{2\mu_i^\rT \Lambda_i^{-1} \tilde{\vz}_i} + \mu_i^\rT \Sigma_i^{-1} \mu_i - \underline{\mu_i^\rT \Lambda_i^{-1} \mu_i} + \underline{2\mu_i^\rT \Lambda_i^{-1} \tilde{\vz}_i} + \underline{\mu_i^\rT \Lambda_i^{-1} \mu_i} 
\end{dmath}
and by noting that the underlined terms cancel out we see that we're left with:
\begin{equation}
\zeta_i = \tilde{\vz}_i^\rT \Gamma_i^{-1} \tilde{\vz}_i + 2\mu_i^\rT \Sigma_i^{-1} \tilde{\vz}_i + \mu_i^\rT \Sigma_i^{-1} \mu_i 
\end{equation}
which is the same as:
\begin{dmath}
\label{Equation:FinalRatio}
\left(\vz_i - \mu_i \right)^\rT \Gamma_i^{-1} \left( \vz_i - \mu_i \right) + 2\mu_i^\rT \Sigma_i^{-1} \left(\vz_i - \mu_i \right) + \mu_i^\rT \Sigma_i^{-1} \mu_i
\end{dmath}
And so $\zeta_i = Q_i$ which completes the proof. \end{proof}
\vspace{2mm}
The key difference between this proof and earlier path integral works which use an application of Girsanov's theorem to sample from a non-zero control input is that this theorem allows for a change in the variance as well. 

In the expression for the likelihood ratio derived here the last two terms ($2\mu_i^\rT \Sigma_i^{-1} \left(\vz_i - \mu_i \right) + \mu_i^\rT \Sigma_i^{-1} \mu_i$) are exactly the terms from Girsanov's theorem. The first term ($\left(\vz_i - \mu_i \right)^\rT \Gamma_i^{-1} \left( \vz_i - \mu_i \right)$), which can be interpreted as penalizing over-aggressive exploration, is the only additional term. 

\subsection{Likelihood Ratio as Additional Running Cost}

The form of the likelihood ratio just derived is easily incorporated into the path integral control framework by folding it into the cost-to-go as an extra running cost. Note that the likelihood ratio appears in both the numerator and denominator of (\ref{Equation:DiscretePI}). Therefore, any terms which do not depend on the state can be factored out of the expectation and canceled. This removes the numerically troublesome normalizing term $\prod_{j=1}^N |A_{t_j}|$. So only the summation of $Q_i$ remains. Recall that $\Sigma = \lambda \vG(\vx_t, t)\vR(\vx_t, t)^{-1}\vG(\vx_t, t)$. This implies that:
\begin{dmath}
\Gamma = \lambda \left( \left( \vG(\vx_t, t)\vR(\vx_t, t)^{-1}\vG(\vx_t, t) \right)^{-1} - \left( A^\rT \vG(\vx_t, t)\vR(\vx_t, t)^{-1}\vG(\vx_t, t)^\rT A \right)^{-1} \right)
\end{dmath}
Now define $\vH = \vG(\vx_t, t)\vR(\vx_t, t)^{-1}\vG(\vx_t, t)^\rT$ and $\tilde{\Gamma} = \frac{1}{\lambda}\Gamma$. We then have:
\begin{dmath}
Q = \frac{1}{\lambda}\left( \left(\vz - \mu \right)^\rT \tilde{\Gamma}^{-1} \left( \vz - \mu \right) + 2\mu^\rT \vH^{-1} \left(\vz - \mu \right) + \mu^\rT \vH^{-1} \mu \right)
\end{dmath}
Then by re-defining the running cost $q(\vx_t, t)$ as:
\begin{dmath}
\tilde{q}(\vx, \vu, \rd \vx) = q(\vx_t, t) + \frac{1}{2}\left(\vz - \mu \right)^\rT \tilde{\Gamma}^{-1} \left( \vz - \mu \right) + \mu^\rT \vH^{-1} \left(\vz - \mu \right) + \frac{1}{2}\mu^\rT \vH^{-1} \mu
\end{dmath}
and $\tilde{S}(\tau) = \phi(\vx_T) + \sum_{j=1}^N \tilde{q}(\vx, \vu, \rd \vx)$, we have:
\begin{equation}
\vu_{t}^* = \vcalG(\vx_t, t)^{-1}\frac{\ExP{\q}{ \exp \left(-\frac{1}{\lambda}\tilde{S}(\tau) \right) \left( \frac{\rd \vx_{t}}{\Delta t} - \vf(\vx_t, t) \right)}}{\ExP{\q}{ \exp \left(-\frac{1}{\lambda}\tilde{S}(\tau) \right)}}
\end{equation}
Also note that $\rd \vx_t$ is now equal to:
\begin{equation}
\left( \vf(\vx_{t}, t) + \vG(\vx_{t}, t) \vu(\vx_t, t) \right) \Delta t + \vB(\vx_t, t)\epsilon\sqrt{\Delta t}
\end{equation}
So we can re-write $\frac{\rd \vx_{t}}{\Delta t} - \vf(\vx_t, t)$ as:
\begin{equation}
\vG(\vx_{t}, t)\vu(\vx_{t}, t) + \vB(\vx_{t}, t)\frac{\epsilon}{\sqrt{\Delta t}}
\end{equation}
And then since $\vG(\vx_{t}, t)$ does not depend on the expectation we can pull it out and get the iterative update law:
\begin{dmath}
\label{Equation:iterativeUpdate}
\vu_{t}^* = \vcalG(\vx_t, t)^{-1} \vG(\vx_t, t)\vu(\vx_t,t) + \vcalG(\vx_t, t)^{-1}\frac{\ExP{\q}{ \exp \left(-\frac{1}{\lambda}\tilde{S}(\tau) \right) \vB(\vx_t, t)\frac{\epsilon}{\sqrt{\Delta t}}}}{\ExP{\q}{ \exp \left(-\frac{1}{\lambda}\tilde{S}(\tau) \right)}}
\end{dmath}

\subsection{Special Case}
The update law (\ref{Equation:iterativeUpdate}) is applicable for a very general class of systems. In this section we examine a special case which we use for all of our experiments. We consider dynamics of the form:
\begin{equation}
\rd \vx_t = \vf(\vx_t, t)\Delta t + \vG(\vx_t ,t)\left(\vu(\vx_t, t)\Delta t + \frac{1}{\sqrt{\rho}}\epsilon\sqrt{\Delta t} \right)
\end{equation}
And for the sampling distribution we set $A$ equal to $\sqrt{\nu}I$. We also assume that $\vG_c(\vx_t, t)$ is a square invertible matrix. This reduces $\vcalH(\vx_t, t)$ to $\vG_c(\vx_t, t)^{-1}$. Next the dynamics can be re-written as:
\begin{equation}
\rd \vx_t = \vf(\vx_t, t)\Delta t + \vG(\vx_t ,t)\left(\vu(\vx_t, t) + \frac{1}{\sqrt{\rho}}\frac{\epsilon}{\sqrt{\Delta t}} \right)\Delta t
\end{equation}
Then we can interpret $\frac{1}{\sqrt{\rho}}\frac{\epsilon}{\sqrt{\Delta t}}$ as a random change in the control input, to emphasize this we will denote this term as $\delta \vu = \frac{1}{\sqrt{\rho}}\frac{\epsilon}{\sqrt{\Delta t}}$. We then have $\vB(\vx_t, t)\frac{\epsilon}{\sqrt{\Delta t}} = \vG(\vx_t, t)\delta \vu$. This yields the iterative update law as:
\begin{dmath}
\label{Equation:ImportanceTerm_new}
\vu(\vx_t, t)^* = \vu(\vx_t, t) + \frac{\ExP{\q}{ \exp \left(-\frac{1}{\lambda}\tilde{S}(\tau) \right) \delta \vu}}{\ExP{\q}{ \exp \left(-\frac{1}{\lambda}\tilde{S}(\tau) \right)}}
\end{dmath}
which can be approximated as:
\begin{equation}
\label{Equation:App_PI}
\vu(\vx_{t_i}, t_i)^* \approx \vu(\vx_{t_i}, t_i) + \frac{\sum_{k=1}^K \exp \left(-\frac{1}{\lambda}\tilde{S}(\tau_{i,k}) \right) \delta \vu_{i,k}}{\sum_{k=1}^K \exp \left(-\frac{1}{\lambda}\tilde{S}(\tau_{i,k}) \right)}
\end{equation}
Where $K$ is the number of random samples (termed rollouts) and $S(\tau_{i,k})$ is the cost-to-go of the $k_{th}$ rollout from time $t_i$ onward. This expression is simply a reward-weighted average of random variations in the control input. Next we investigate what the likelihood ratio addition to the running cost is. For these dynamics we have the following simplifications:
\begin{enumerate}
\item $\vz - \mu = \vG(\vx_t, t)\delta \vu$
\item $\tilde{\Gamma}^{-1} = (1 - \nu^{-1})\vG(\vx_t, t)^{-1}\vR(\vx_t, t)\vG(\vx_t, t)$
\item $\vH^{-1} = \vG(\vx_t, t)^{-1}\vR(\vx_t,t)\vG(\vx_t, t)^{-1}$
\end{enumerate}
Given these simplifications $\tilde{q}$ reduces to:
\begin{dmath}
\tilde{q}(\vx, \vu, \rd \vx) = q(\vx_t, t) + \frac{(1 - \nu^{-1})}{2}\delta \vu^\rT \vR \delta \vu + \vu^\rT \vR \delta \vu + \frac{1}{2}\vu^\rT \vR \vu
\end{dmath}
This means that the introduction of the likelihood ratio simply introduces the original control cost from the optimal control formulation into the sampling cost, which originally only included state-dependent terms.

\section{MODEL PREDICTIVE CONTROL ALGORITHM}

\begin{figure}[b]
  \centering
  \includegraphics[width=.9\columnwidth, trim = 0mm 0mm 0mm 10mm, clip=true]{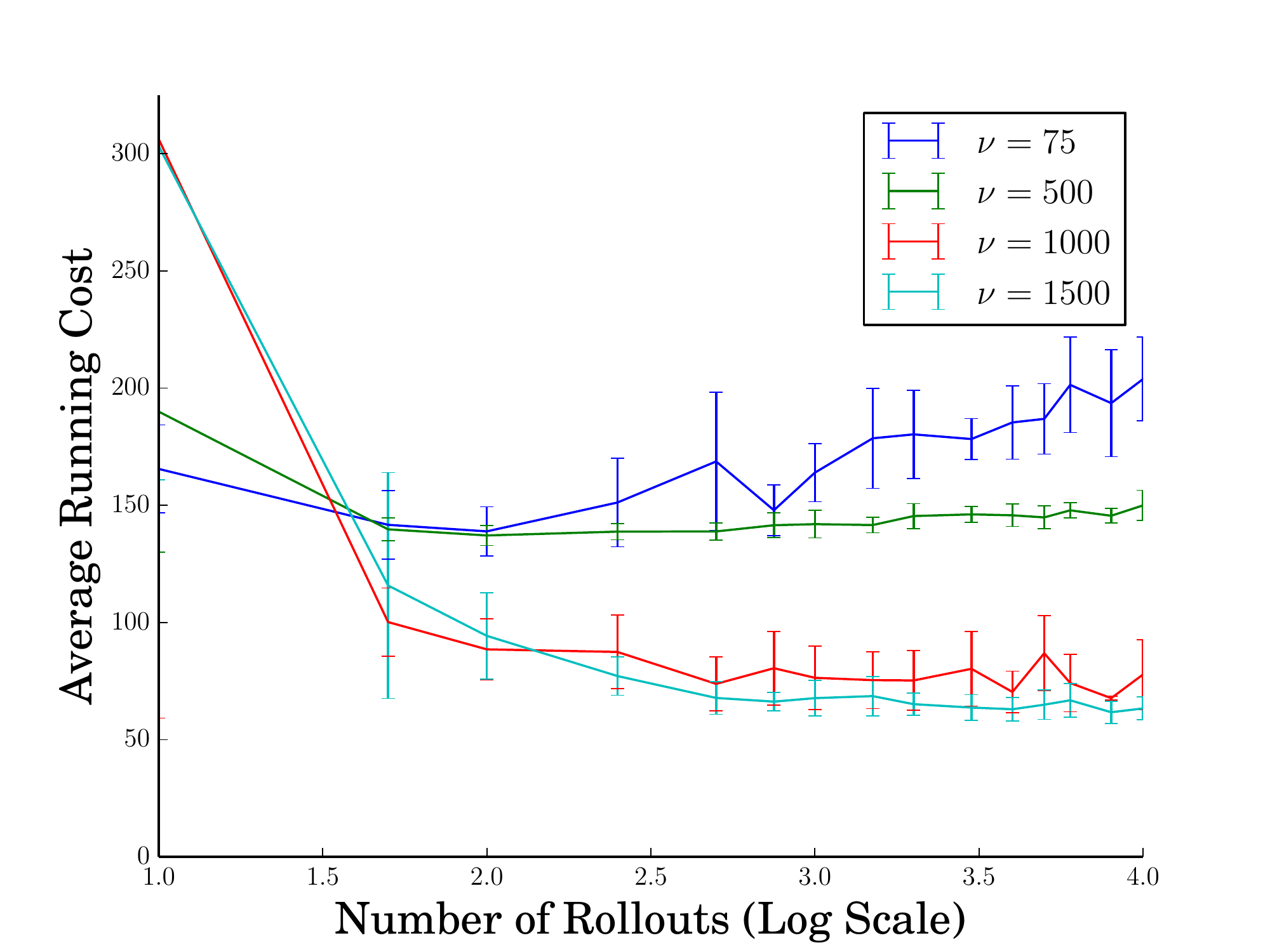}
  \caption{Average running cost for the cart-pole swing-up task as a function of the exploration variance $\nu$ and the number 
  	       of rollouts. Using only the natural system variance the MPC algorithm does not converge in this scenario.}
\label{fig:Cart_Pole_Cost_Comparison}
\end{figure}

We apply the iterative path integral control update law, with the generalized importance sampling term, in a model predictive control setting. In this setting optimization and execution occur simultaneously: the trajectory is optimized and then a single control is executed, then the trajectory is re-optimized using the un-executed portion of the previous trajectory to warm-start the optimization. This scheme has two key requirements:
\begin{enumerate}
\item Rapid convergence to a good control input.
\item The ability to sample a large number of trajectories in real-time.
\end{enumerate}
The first requirement is essential because the algorithm does not have the luxury of waiting until the trajectory has converged before executing. The new importance sampling term enables tuning of the exploration variance which allows for rapid convergence, this is demonstrated in Fig. \ref{fig:Cart_Pole_Cost_Comparison}. 

The second requirement, sampling a large number of trajectories in real-time, is satisfied by implementing the random sampling of trajectories on a GPU. The algorithm is given in Algorithm \ref{Algorithm:mppi}, in the parallel GPU implementation the sampling for loop (for k to K-1) is run completely in parallel.

\begin{algorithm}
\SetKwInOut{Input}{Given}
\Input{$K$: Number of samples\;
       $N$: Number of timesteps\;
       $(\vu_0, \vu_1, ... \vu_{N-1})$: Initial control sequence\;
       $\Delta t , \vx_{t_0}, \vf, \vG, \vB, \nu$: System/sampling dynamics\;
       $\phi, q, \vR, \lambda$: Cost parameters\; 
       $\vu_{\text{init}}$: Value to initialize new controls to\;}
        
\BlankLine

\While{task not completed}{

\For{$k \leftarrow 0$ \KwTo $K-1$}{
  $\vx = \vx_{t_0}$\;
  \For{$i \leftarrow 1$ \KwTo $N-1$}{
    $\vx_{i+1} = \vx_{i} + \left( \vf + \vG \left( \vu_{i} + \delta \vu_{i,k} \right) \right) \Delta t$\; 
  	$\tilde{S}(\tau_{i+1, k}) = \tilde{S}(\tau_{i,k}) + \tilde{q}$\;
  	}
}
\BlankLine
\For{$i \leftarrow 0$ \KwTo $N-1$}{
$\vu_i \leftarrow \vu_i + \left[ \sum_{k=1}^K \left( \frac{\exp \left( -\frac{1}{\lambda} \tilde{S}_(\tau_{i,k}) \right) \delta \vu_{i,k}}{\sum_{k=1}^K \exp \left( -\frac{1}{\lambda} \tilde{S}_(\tau_{i,k}) \right)} \right) \right]$\;
}

$\text{send to actuators}(\vu_0)$\;

\For{$i \leftarrow 0$ \KwTo $N-2$}{ 
	$\vu_i = \vu_{i+1}$\;  
}
$\vu_{N-1} = \vu_{\text{init}}$
  
Update the current state after receiving feedback\; 
check for task completion\;
}

\caption{Model Predictive Path Integral Control \label{Algorithm:mppi}}
\end{algorithm}

\section{EXPERIMENTS}

We tested the model predictive path integral control algorithm (MPPI) on three simulated platforms (1) A cart-pole, (2) A miniature race car, and (3) A quadrotor attempting to navigate an obstacle filled environment. For the race car and quadrotor we used a model predictive control version of the differential dynamic programming (DDP) algorithm as a baseline comparision. In all of these experiments the controller operates at 50 Hz, this means that the open loop control sequence is re-optimized every 20 milliseconds.
\subsection{Cart-Pole} For the cart-pole swing-up task we used the state cost: $q(\vx) = p^2 + 500(1 + \cos(\theta))^2 + \dot{\theta}^2 + \dot{p}^2$, where $ p $  is the position of cart, $ \dot{p} $ is the velocity and $ \theta,\dot{\theta} $  are the angle and angular velocity of the pole. The control input is desired velocity, which maps to velocity through the equation: $\ddot{p} = 10(u - \dot{p})$. The disturbance parameter $\frac{1}{\sqrt{\rho}}$ was set equal $.01$ and the control cost was $\vR = 1$.  We ran the
MPPI controller for 10 seconds with a 1 second optimization horizon. The controller has to swing-up the pole and keep it balanced for the rest of the 10 second horizon. The exploration variance parameter, $\nu$, was varied between $1$ and $1500$. The MPPI controller is able to swing-up the pole faster with increasing exploration variance. Fig. \ref{fig:Cart_Pole_Cost_Comparison} illustrates the performance of the MPPI controller as the exploration variance and the number of rollouts are changed. Using only the natural variance of the system for exploration is insufficient in this task, in that case (not shown in the figure) the controller is never able to swing-up the pole which results in a cost around 2000.

\subsection{Race Car} 
\begin{figure}[ht!]
\centering
\includegraphics[width=\columnwidth, trim = 20mm 3mm 20mm 5mm]{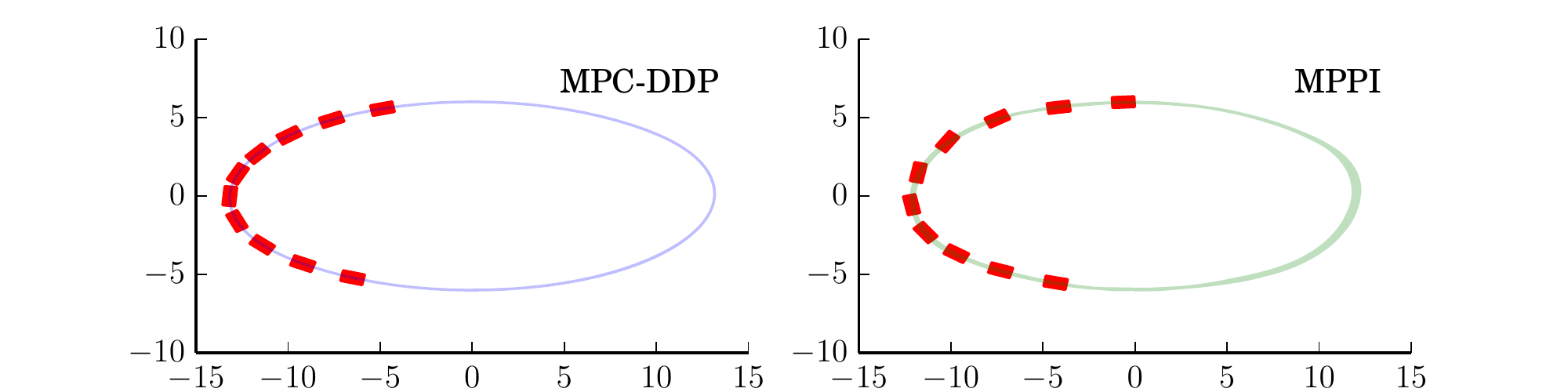}
\caption{Comparison of DDP (left) and MPPI (right) performing a cornering maneuver along an ellipsoid track. MPPI is able
         to make a much tigther turn while carrying more speed in and out of the corner than DDP. The direction of travel is counterclockwise.}
\label{fig:race_car_traj}
\end{figure}

\begin{figure}[ht!]
\centering
\includegraphics[width=.9\columnwidth, trim = 10mm 10mm 10mm 10mm]{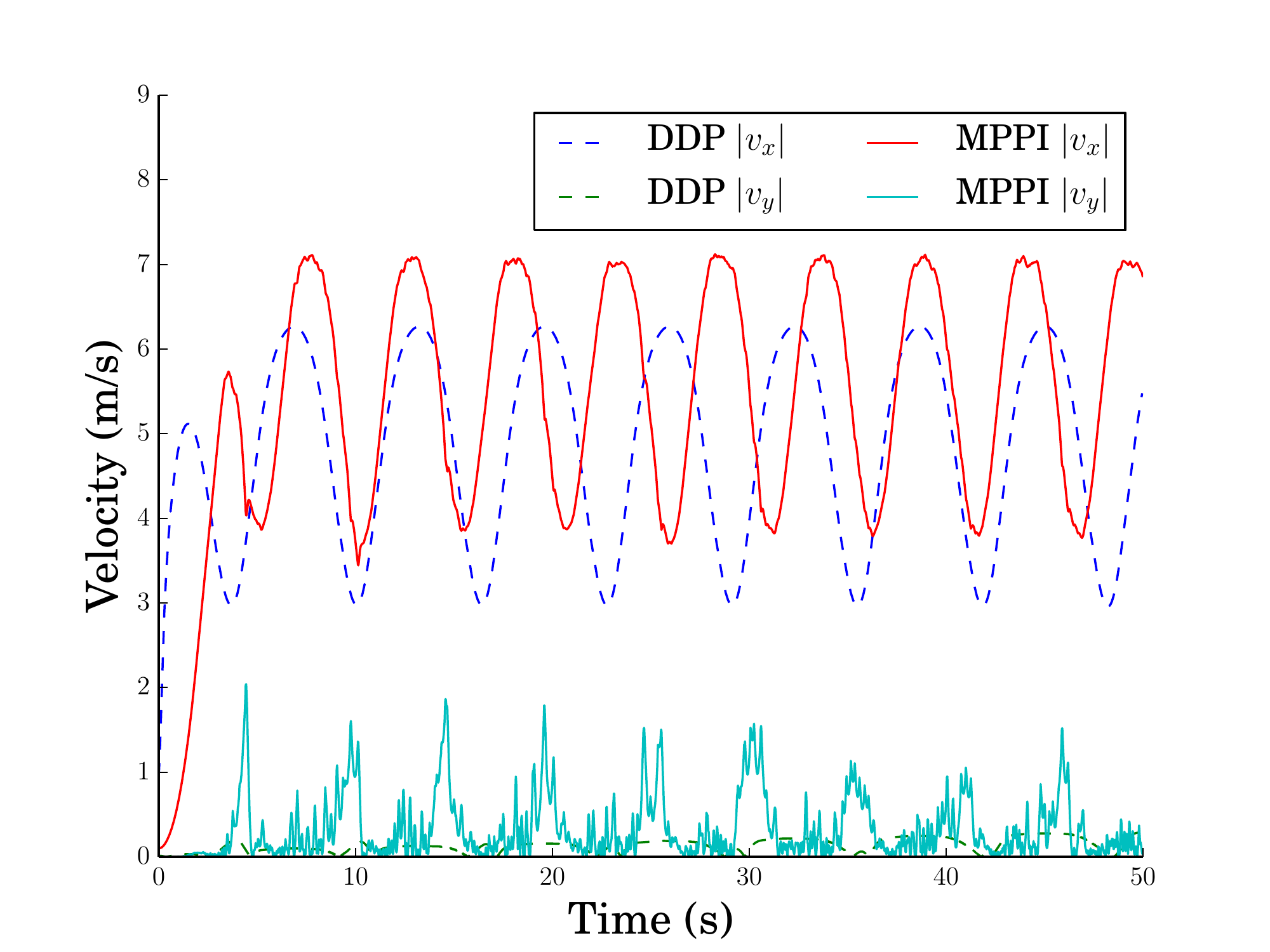}
\caption{Comparison of DDP (left) and MPPI (right) performing a cornering maneuver along an ellipsoid track. MPPI is able
         to make a much tigther turn while carrying more speed in and out of the corner than DDP.}
\label{fig:race_car_stats}
\end{figure}

In the race car task the goal was to minimize the objective function: $q(\vx)  = 100d^2 + (v_x - 7.0)^2$. Where $d$ is defined as: $d = |\left(\frac{x}{13}\right)^2 + \left(\frac{y}{6}\right)^2 - 1|$, and $v_x$ is the forward (in body frame) velocity of the car. This cost ensures that the car to stays on an elliptical track while maintaining a forward speed of 7 meters/sec. We use a non-linear dynamics model \cite{HindThesis}  which takes into account the (highly non-linear) interactions between tires and the ground. The exploration variance was set to a constant $\nu$ times the natural variance of the system.
\begin{figure}[ht!]
\centering
\includegraphics[width=.9\columnwidth, trim = 0mm 2mm 0mm 10mm]{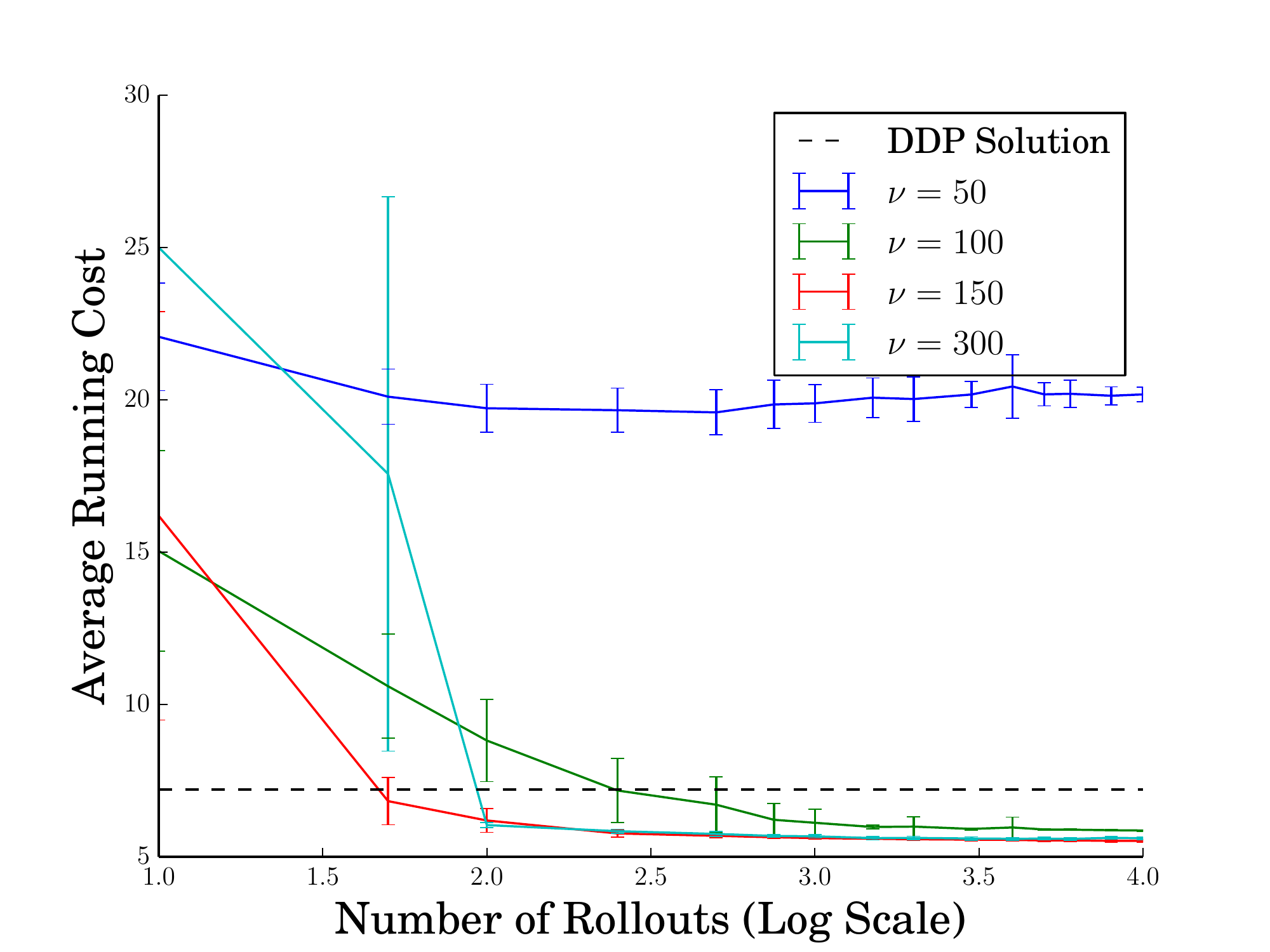}
\caption{Performance comparison in terms of average cost between  MPPI and MPC-DDP as the exploration variance $\nu$ changes from 50 to 300 and the
         number of rollouts changes from 10 to 1000. Only with a very large increase in the exploration variance is MPPI able to outperform MPC-DDP. Note that the cost is capped at 25.0}\label{fig:}
\label{fig:Race_Car_Cost}
\end{figure}
The MPPI controller is able to enter turns at close to the desired speed of 7 m/s and then slide through the turn. The DDP solution does not attempt to slide and significantly reduces its forward velocity before entering the turn, this results in a higher average cost compared to the MPPI controller. Fig. \ref{fig:Race_Car_Cost} shows the cost comparison between MPPI and MPC-DDP, and Figures \ref{fig:race_car_traj} and \ref{fig:race_car_stats} show samples of the trajectories taken by the two algorithms as well as the velocity profiles.

\subsection{Quadrotor} The quadrotor task was to fly through a field filled with cylindrical obstacles as fast as possible. We used the quadrotor dynamics model from \cite{michael2010grasp}. This is a non-linear model which includes position, velocity, euler angles, angular acceleration, and the rotor dynamics. We randomly generated three forests, one where obstacles are on average 3 meters apart, the second one 4 meters apart, and the third 5 meters apart. We then separately created cost functions for both MPPI and DDP which guide the quadrotor through the forest as quickly as possible.
\begin{figure}[ht!]
\centering
\includegraphics[width=\columnwidth, trim = 10mm 5mm 10mm 8mm, clip=true]{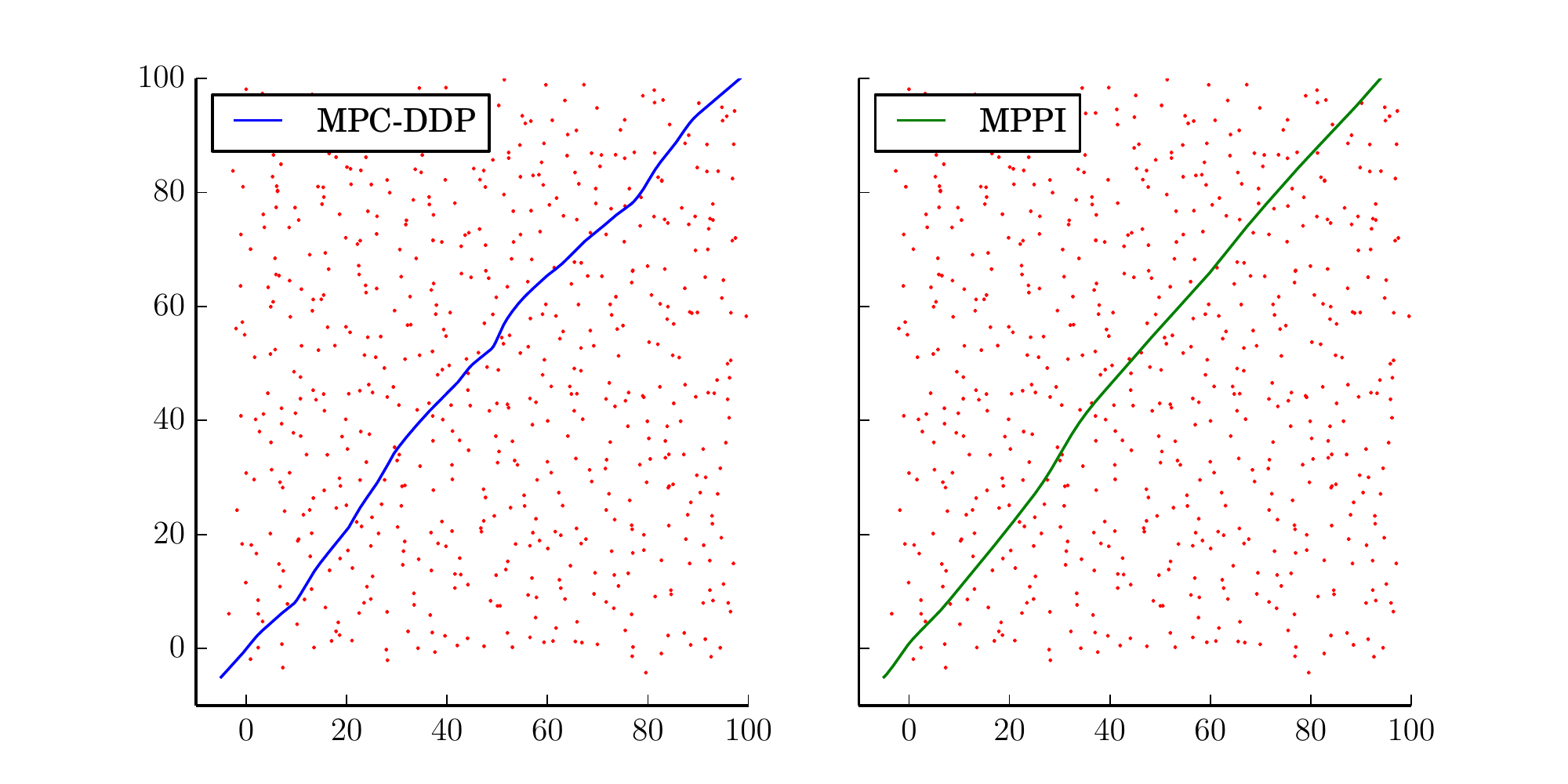}
\caption{Left: sample DDP trajectory through 4m obstacle field, Right: Sample MPPI trajectory through the same field. Since the MPPI controller can directly reason
about the shape of the obstacles it is able to safely pass through the field taking a much more direct route.}
\label{quad_race}
\end{figure}
The cost function for MPPI was of the form:
$q(\vx) = 2.5(p_x - p_x^{des})^2 + 2.5(p_y - p_y^{des})^2 + 150(p_z - p_z^{des})^2 + 50 \psi^2  + \|v\|^2 + 350\exp(-\frac{d}{12}) + 1000C$
where $(p_x, p_y, p_z)$ denotes the position of the vehicle. $\psi$ denotes the yaw angle in radians, v is velocity, and $d$ is the distance to the closest obstacle.
$C$ is a variable which indicates whether the vehicle has crashed into the ground or an obstacle. Additionally if $C = 1$ (which indicates a crash), the rollout stops simulating the dynamics and the vehicle remains where it is for the rest of the time horizon. We found that the crash indicator term is not useful for the MPC-DDP based controller, this is not surprising since the discontinuity it creates is difficult to approximate with a quadratic function. The term in the cost for avoiding obstacles in the MPC-DDP controller consists purely of a large exponential term: $2000 \sum_{i=1}^N \exp (-\frac{1}{2} d_i^2 )$, note that this sum is over all the obstacles in the proximity of the vehicle whereas the MPPI controller only has to consider the closest obstacle. 

\begin{figure}[ht!]
\centering
\includegraphics[width = .9\columnwidth, trim = 5mm -4mm 5mm 10mm, clip=true]{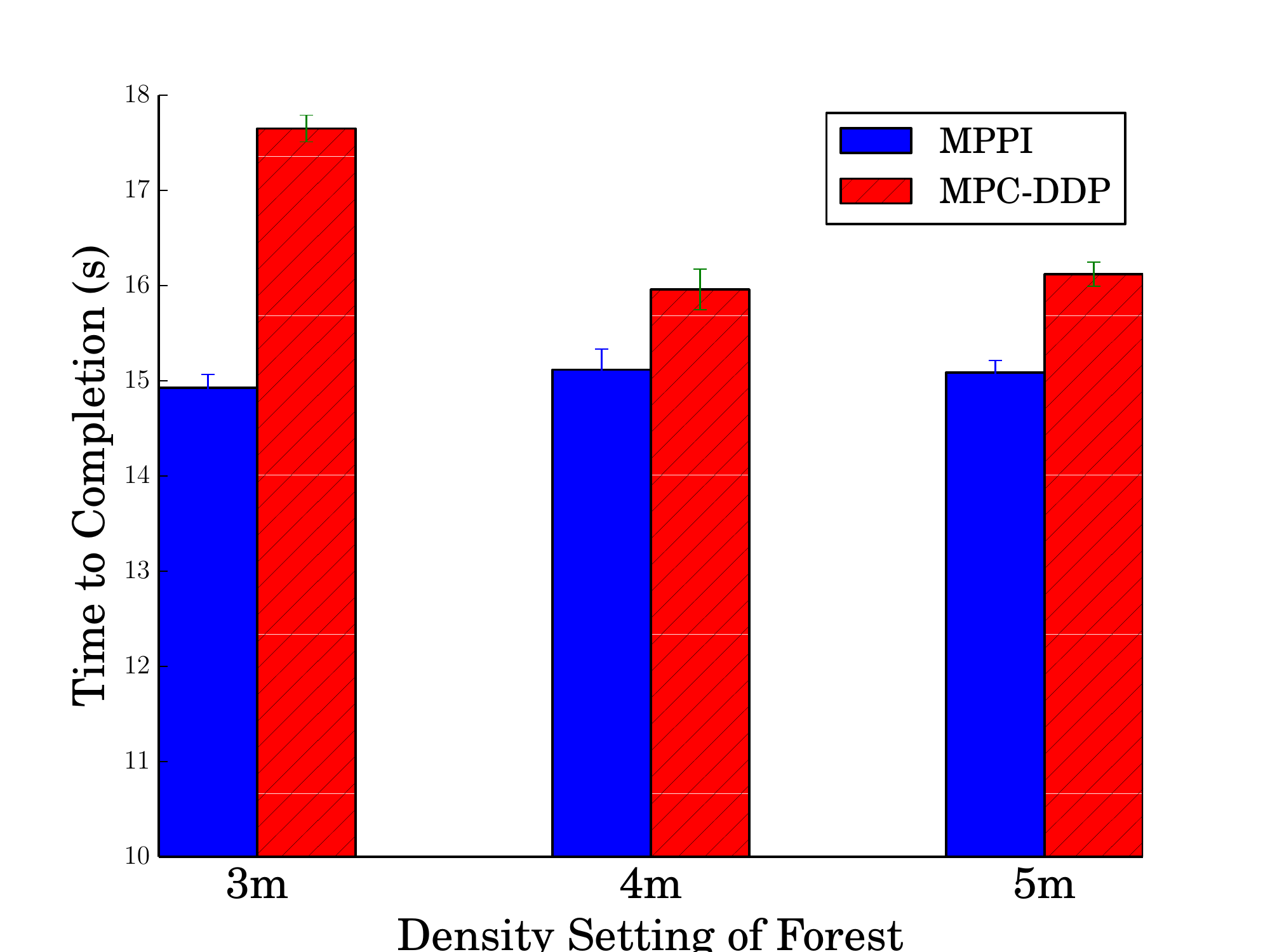}
\caption{Time to navigate forest. Comparison between MMPI and DDP.}
\label{fig:quad_race}
\end{figure}

\begin{figure}[ht!]
\centering
\includegraphics[width = \columnwidth, trim = 5mm 50mm 5mm 40mm, clip=true]{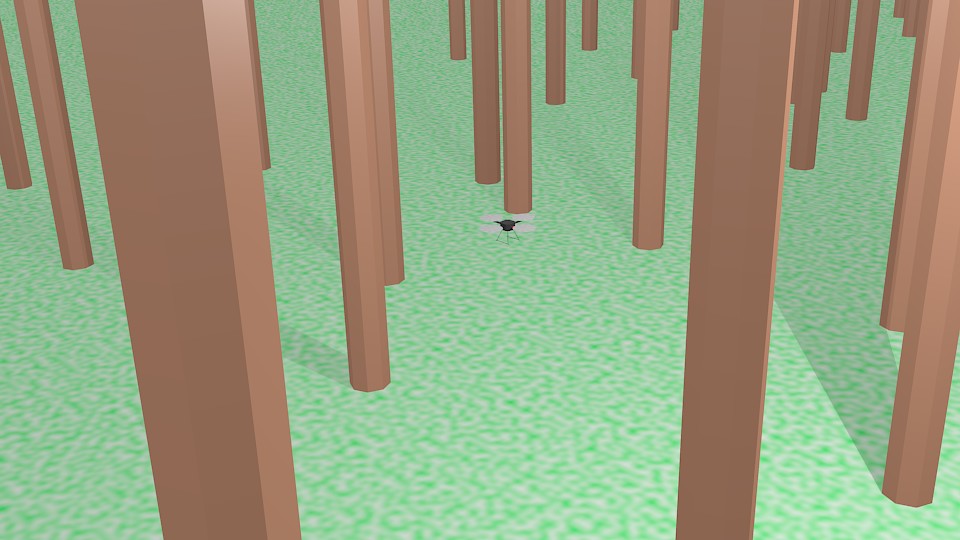}
\caption{Simulated forest environment used in the quadrotor navigation task.}
\label{fig:quad_time}
\end{figure}

Since the MPPI controller can explicitly reason about crashing (as opposed to just staying away from obstacles), it is able to travel both faster and closer to obstacles than the MPC-DDP controller. Fig. \ref{fig:quad_time} shows the difference in time between the two algorithms and Fig. \ref{fig:quad_race} the trajectories taken by MPC-DDP and one of the MPPI runs on the forest with obstacles placed on average 4 meters away.

\section{CONCLUSION}
In this paper we have developed a model predictive path integral control algorithm which is able to outperform a state-of-the-art DDP method on two difficult control tasks. The algorithm is based on stochastic sampling of system trajectories and requires no derivatives of either the dynamics or costs of the system. This enables the algorithm to naturally take into account non-linear dynamics, such as a non-linear tire model \cite{HindThesis}. It is also able to handle cost functions which are intuitively appealing, such as an impulse cost for hitting an obstacle, but are difficult for traditional approaches that rely on a smooth gradient signal to perform optimization. The two keys to achieving this level of performance with a sampling based method are:
\begin{enumerate}
\item The derivation of the generalized likelihood ratio between discrete time diffusion processes.
\item The use of a GPU to sample thousands of trajectories in real-time.
\end{enumerate}
The derivation of the likelihood ratio enables the designer of the algorithm to tune the exploration variance in the path integral control framework, whereas previous methods have only allowed for the mean of the distribution to be changed. Tuning the exploration variance is critical in achieving a high level of performance since the natural variance of the system is typically too low to achieve good performance.

The experiments considered in this work only consider changing the variance by a constant multiple times the natural variance of the system. In this special case the introduction of the likelihood ratio corresponds to adding in a control cost when evaluating the cost-to-go of a trajectory. A direction for future research is to investigate how to automatically adjust the variance online. Doing so could enable the algorithm to switch from aggressively exploring the state space when performing aggressive maneuvers to exploring more conservatively for performing very precise maneuvers.   

\bibliographystyle{IEEEtran}
\bibliography{References}

\end{document}